\documentclass[final,sort&compress]{aipproc}

\layoutstyle{6x9}

\begin{document}

\title{Confinement and center vortex dynamics in different gauge groups}


\author{M.~Engelhardt}{
  address={Department of Physics, New Mexico State University,
           Las Cruces, NM 88003, USA}
}

\author{B.~Sperisen}{
  address={Department of Physics, New Mexico State University,
           Las Cruces, NM 88003, USA}
}

\begin{abstract}
The random vortex world-surface model is extended to the gauge groups
$SU(4)$ and $Sp(2)$. Compared to the $SU(2)$ and $SU(3)$ models studied
previously, which reproduce the infrared properties of the corresponding
Yang-Mills theories on the basis of a simple vortex world-surface curvature
action, new dynamical characteristics become important. In the $SU(4)$ case,
an explicit dependence of the vortex effective action on the configuration
of the Abelian magnetic monopoles residing on the vortices emerges; in
the $Sp(2)$ case, a new ``stickiness'' contribution to the vortex action
serves to drive the deconfinement phase transition towards the correct
first-order behavior.
\end{abstract}

\maketitle

\section{Introduction}
The random vortex world-surface model is a concrete realization of the
center vortex picture of the strong interaction vacuum
\cite{hooft,olesen,jg2,tk1,df1,per,jg3}, i.e., the notion that the
relevant infrared gluonic degrees of freedom of the strong interaction
are closed tubes of quantized chromomagnetic flux. The random vortex
world-surface model was initially investigated for $SU(2)$
Yang-Mills theory \cite{m1,m2,m3}, and in this simplest case, the
main characteristics of the strongly interacting vacuum were reproduced.
Both a confining low-temperature phase as well as a deconfined
high-temperature phase are found \cite{m1}, separated by a second-order
deconfinement phase transition; furthermore, the topological susceptibility
\cite{m2,cw2,contvort,bruck} and the (quenched) chiral condensate
\cite{m3} match the ones extracted from $SU(2)$ lattice Yang-Mills theory
quantitatively. Extending the investigation to the $SU(3)$ gauge group
\cite{su3conf,su3bary,su3freee}, the deconfinement phase transition
exhibits weakly first-order behavior \cite{su3conf}, and a Y-law for the
baryonic static potential is found \cite{su3bary}, again matching the
corresponding characteristics of $SU(3)$ lattice Yang-Mills theory.
Studies of the topological and chiral properties in the $SU(3)$ case
are pending.

The aforementioned successes of the $SU(2)$ and $SU(3)$ random vortex
world-surface models are obtained on the basis of very simple vortex
dynamics, with the action determined purely by world-surface curvature.
Accordingly, these models only contain one dimensionless coupling
parameter, which is adjusted in practice to reproduce the ratio of the
deconfinement temperature $T_c $ to the square root of the zero-temperature
string tension $\sigma $ found in the corresponding Yang-Mills theory.
Recent efforts have focused on the question of how far this simple
picture carries as the gauge group is varied. There are two systematic
ways of extending the Yang-Mills gauge group beyond the cases discussed
above: On the one hand, one may increase the number of colors $N$
determining the $SU(N)$ gauge symmetry; on the other hand
\cite{spn,g21o,pepelat05}, the $SU(2)$ group can alternatively be
considered as the smallest symplectic group $Sp(1)$, and the $Sp(N)$
sequence can also be used to generalize $SU(2)=Sp(1)$. Accordingly,
random vortex world-surface models for the infrared sectors of both
$SU(4)$ and $Sp(2)$ Yang-Mills theory have been constructed
\cite{su4,sp2full} and are presented in the following. In both cases,
new dynamical characteristics emerge.

The concrete modeling methodology used in the random vortex world-surface
model is discussed in detail in \cite{m1,su3conf,su4}. Vortex world-surfaces
are composed of elementary squares on a hypercubic space-time lattice.
The lattice spacing is a fixed physical quantity related to the transverse
thickness of vortices; it represents the ultraviolet cutoff inherent in
any infrared effective description. An ensemble of closed vortex
world-surfaces is generated by Monte Carlo update. For different
underlying gauge groups, random vortex world-surface models differ in two
respects: On the one hand, the quantization of vortex flux is determined
by the center of the gauge group. When encircling a vortex, a Wilson loop
acquires a phase given by one of the nontrivial center elements.
Accordingly, in general, several species of center vortices, corresponding
to the different nontrivial center elements, can exist. They can merge and
disassociate into one another. On the other hand, the models can differ in
the effective action governing the vortices.

\section{$SU(4)$ vortex model}
The $SU(4)$ group contains the nontrivial center elements $\{ i,-i,-1\} $.
Vortex fluxes associated with the elements $i$ and $-i$ are related by an
inversion of space-time orientation; therefore, there are altogether only
two physically distinct types of center vortices. A vortex generating a
phase factor $-1$ when linked to a Wilson loop can branch into two vortices
associated with a phase factor $\pm i$ and vice versa.

Correspondingly, $SU(4)$ Yang-Mills theory \cite{luctep03,luctep04,luctepnew}
also induces two distinct string tensions, the quark string tension
$\sigma_{1} $ and the diquark string tension $\sigma_{2} $. The $SU(4)$
Yang-Mills confinement properties thus are characterized by the ratios
$\sigma_{2} /\sigma_{1}$ and $T_c / \sqrt{\sigma_{1} } $, as well as
the behavior at the deconfinement phase transition, which is about twice
as strongly first-order as the one of $SU(3)$ Yang-Mills theory.

To properly model these characteristics, it is necessary to use an effective
vortex action which is more complicated than for $SU(2)$ or $SU(3)$,
and which can be symbolically represented as
\vspace{0.3cm}
\begin{equation}
\hspace{-0.4cm} S\ \ = \ \ c_i \, \, c_j \ \times
\hspace{1.5cm} - \ \ b \ \times
\ \ \ \ \ \ \ \ \ \ \ \ \ \ \ \ .
\label{su4voract}
\end{equation}
\vspace{-1.8cm}

\begin{figure}[h]
\centerline{\hspace{1.6cm} \includegraphics[width=1.7cm]{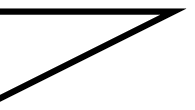}
\hspace{0.6cm} \includegraphics[width=1.53cm]{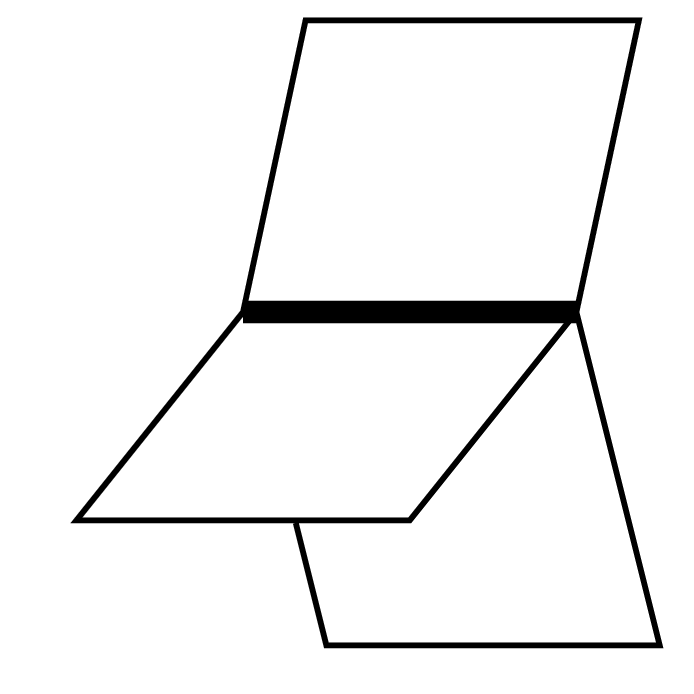} }
\end{figure}

\vspace{-2.1cm}

\hspace{6.4cm} $j$

\vspace{1cm}

\hspace{6.05cm} $i$

The first term is the curvature term already used in the $SU(2)$ and the
$SU(3)$ models. It penalizes configurations in which two vortex squares
share a link without lying in the same plane by an action increment
$c_i c_j $ depending on the types of vortices participating; in the
$SU(4)$ case, there are two vortex types, $i,j \in \{ 1,2\} $. Studying
a model based only on this term led to the conclusion that it cannot
faithfully reproduce the confinement properties of $SU(4)$. When
$\sigma_{2} /\sigma_{1} $ is tuned to the correct value,
the deconfinement phase transition is second-order. Consequently,
additional dynamics must be introduced, embodied in the second term in
\eqref{su4voract}, the branching term. It facilitates vortex branchings
by weighting links at which 3 or 5 vortex squares meet with an
action decrement $b$.

Using this action, agreement with the $SU(4)$ Yang-Mills confinement
characteristics is reached at the physical point \cite{su4}
\begin{equation}
c_1 = 0.45 \ \ \ \ \ \ \ \ \ \ \ \ c_2 = 0.80
\ \ \ \ \ \ \ \ \ \ \ \ b=0.71 \ .
\label{physparms}
\end{equation}
It should be noted that, in Abelian gauges, vortex branching can be
associated with Abelian magnetic monopoles \cite{su4}. Thus, the above result
can be interpreted as implying that a realistic vortex model for $SU(4)$
Yang-Mills theory is only achieved by including a dependence of the
vortex dynamics on the configuration of the Abelian magnetic monopoles which
reside on the vortices in Abelian gauges \cite{m2,cw2,contvort}.
This confirms a corresponding expectation formulated in \cite{jeffstef},
that Abelian magnetic monopoles begin to play a role in infrared Yang-Mills
vortex dynamics as the number of colors $N$ is raised. Note that Abelian
magnetic monopoles are also present in $SU(2)$ and $SU(3)$ vortex
configurations; however, there is no signature for an independent
dynamical role of these monopoles. Their distribution appears to be
essentially determined by the dynamics of the vortices on which they reside.

\section{$Sp(2)$ vortex model}
A remarkable property of the $Sp(N)$ sequence of groups is that all members
have the same center, $Z(2)$, and allow for the same set of center vortex
degrees of freedom. There is only one nontrivial center element, $-1$, and
therefore only one type of vortex flux. Nevertheless, the effective vortex
actions can be very different for different underlying $Sp(N)$ groups;
after all, different cosets would be integrated out in each case if one
were to derive the effective vortex action from first principles.
Therefore, vortex models for different $Sp(N)$ Yang-Mills theories are
by no means forced to display similar confinement characteristics, such
as deconfinement transitions of the same order.

Indeed, while $SU(2)=Sp(1)$ Yang-Mills theory exhibits a second-order
deconfinement phase transition, the deconfinement transition of $Sp(2)$ 
Yang-Mills theory is strongly first-order \cite{spn,g21o}. As above, in
order to generate such behavior, new dynamics must be introduced
compared to the $SU(2)$ vortex model. The confinement characteristics
of $Sp(2)$ Yang-Mills theory can be reproduced using an effective vortex
action of the symbolic form
\vspace{-0.15cm}

\begin{equation}
S\ \ = \ \ c \ \times
\hspace{1.9cm} + \ \ \ \ s \ \times
\hspace{2cm} .
\label{voract}
\end{equation}
\vspace{-1.9cm}

\begin{figure}[h]
\centerline{\hspace{1.4cm} \includegraphics[width=1.7cm]{cactdef.ps}
\hspace{1.3cm} \includegraphics[width=1.53cm]{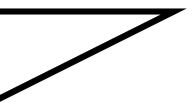} }
\end{figure}
\vspace{-0.3cm}

The first term is the curvature term already discussed above. The
second term can be interpreted in terms of a ``stickiness'' of vortices:
When 4 (or even 6) vortex squares meet at a link, this corresponds to
2 (or even 3) intersecting vortex fluxes maintaining contact to one another
for a finite space-time length instead of intersecting only at one space-time
point. Enhancing such behavior by choosing a negative value for $s$ means
that vortices become stickier. Indeed, a first-order deconfinement phase
transition of the proper strength, together with the correct value of
$T_c / \sqrt{\sigma } $ is achieved at the physical point \cite{sp2full}
\begin{equation}
c = 0.479 \ \ \ \ \ \ \ \ \ \ \ \ \ \ \ s = -1.745 \ .
\label{physcs}
\end{equation}

\section{Conclusions}
Extending the Yang-Mills gauge group to $SU(4)$ and $Sp(2)$, new dynamics
emerge in the corresponding infrared effective vortex descriptions. The
$SU(4)$ case exhibits clear signatures of Abelian magnetic monopoles
(which are intrinsically present in vortex configurations cast in Abelian
gauges) attaining a dynamical significance of their own as the number of
colors is raised. This corroborates related arguments put forward in
\cite{jeffstef}. In the $Sp(2)$ case, a new ``stickiness'' term in
the effective action serves to drive the deconfinement transition towards
the correct first-order behavior. While $SU(2)=Sp(1)$ and $Sp(2)$ Yang-Mills
theory contain the same center vortex degrees of freedom, the vortex
effective actions in the two cases differ and thus naturally lead to
different behavior at the deconfinement transition.

Having determined the physical points \eqref{physparms} and \eqref{physcs}
of the $SU(4)$ and $Sp(2)$ infrared effective vortex models, the behavior
of the spatial string tensions at high temperatures can be predicted
\cite{su4,sp2full}. As discussed further in \cite{su4,sp2full}, comparison
with measurements in the corresponding full lattice Yang-Mills theories can
be used to test the validity of the model constructions presented here.
\vspace{-0.15cm}

\begin{theacknowledgments}
\vspace{-0.1cm}
This work was supported by the U.S.~DOE under grants DE-FG03-95ER40965
(M.E.) and DE-FG02-94ER40847 (B.S.).
\end{theacknowledgments}

\end{document}